\documentclass[11pt]{article}
\usepackage{jcappub}
\usepackage{natbib,float,caption,comment,bm}
\title{Mitigating the impact of fiber assignment on the measurement of galaxy-lensing cross correlation}
\author[a,b]{Ryu Makiya}
\author[c]{and Tomomi Sunayama}
\affiliation[a]{Institute of Astronomy and Astrophysics, Academia Sinica, Astronomy-Mathematics Building, AS/NTU, No. 1, Section 4, Roosevelt Road, Taipei 10617, Taiwan}
\affiliation[b]{Kavli Institute for the Physics and Mathematics of the
Universe, Todai Institutes for Advanced Study, the University of Tokyo,
Kashiwa, Japan 277-8583 (Kavli IPMU, WPI)}
\affiliation[c]{Department of Physics, Nagoya University, Furo-cho, Chikusa-ku, Nagoya, Aichi, Japan 464-8601}
\emailAdd{rmakiya@asiaa.sinica.edu.tw,sunayama.tomomi@d.mbox.nagoya-u.ac.jp}

\abstract{
We examine the impact of fiber assignment on the measurement of galaxy clustering and its cross correlation with weak lensing fields.
Unlike the past spectroscopic galaxy surveys such as Baryon Oscillation Spectroscopic Survey (BOSS), currently ongoing spectroscopic galaxy surveys such as Prime Focus Spectrograph (PFS) and Dark Energy Spectroscopic Instrument (DESI) suffer from the fiber assignment artifacts more severely because there are more target galaxies than available fibers.
The previous studies found that the fiber assignment suppresses the amplitude of the galaxy power spectrum at all scales.
We newly find that the fiber assignment introduces the artificial correlation of structure at different redshifts, which suppresses the amplitude of the galaxy-lensing cross power spectrum.
We show that the fiber assignment effects on the cross power spectrum can be mitigated at all scales with accuracy better than $\sim 1\%$, by up-weighting observed galaxies with the probability to be observed.
This is not the case for the galaxy power spectrum, which is not fully corrected at $k \gtrsim 0.2$ [$h$/Mpc].
We find that the galaxy-lensing cross power spectrum is not affected by the pairwise probability of galaxies to be observed, and thus the correction method based on the individual probability is sufficient at all scales. 
}

\begin{document}
\maketitle
\flushbottom

\section{Introduction}
Spectroscopic galaxy surveys yield three-dimensional distributions of galaxies in redshift space, which enable us to constrain the cosmological models through the measurement of, e.g., baryon acoustic oscillation (BAO) and redshift space distortion \cite{eBOSS:2020}.
On-going and near-future spectroscopic galaxy surveys such as the Subaru Prime Focus Spectrograph (PFS) \cite{pfs_takada}, the Dark Energy Survey Instrument (DESI) \cite{DESI,DESI2}, HETDEX \cite{HETDEX}, Euclid \cite{EUCLID}, {\it Roman} space telescope \cite{WFIRST}, and Legacy Survey of Space and Time by Vera C. Rubin Observatory\cite{LSST}, as well as future surveys such as MegaMapper \cite{MegaMapper}, will continue to expand the survey volume with higher galaxy number density.
The large imaging surveys such as the Subaru Hyper Suprime Cam (HSC) \cite{HSC1,HSC2},  Kilo-Degree Survey (KiDS) \cite{kids}, and Dark Energy Survey (DES) \cite{DES1,DES2} provide complementary information through the weak-lensing observation and its cross-correlation with spectroscopic galaxies.
These surveys will dramatically increase the statistical precision of cosmological parameters, which in turn requires more careful handling of systematic uncertainties.

In most spectroscopic galaxy surveys, the fiber-fed multi-object spectrograph system is used to maximize the redshift accuracy.
In these systems, galaxy spectra are taken by placing the optical fibers, which are connected to the spectrograph, on target galaxies.
The spectroscopic survey can only observe a fraction of the parent targets due to finite observing times, which results in the complicated spatial pattern of missing objects and the non-trivial bias in the observed galaxy clustering.
One example is the so-called ``fiber collision'' effect: two objects which are too close to place two fibers cannot be observed and the angular clustering at the scale smaller than the physical size of fibers is suppressed.

One approach taken in the Sloan Digital Sky Survey (SDSS) is to simply assign the same redshift to the missing objects with the nearest target object \cite[e.g.,][]{SDSS:2004oes,SDSS:2006ymz}, assuming that these galaxies are likely to be physically associated.
This standard approach is so-called the``nearest neighbor'' method \cite{anderson/etal:2012}.
Another approach is to weight observed galaxies using the angular correlation function of the parent photometric catalog \cite{hawkins/etal:2003,li/etal:2006,white/etal:2011,okumura/etal:2016}. 
Several other methods to correct for missing targets are proposed in the literature.
For example, ref.\cite{burden/etal:2017,pinol/etal:2017} reduce the fiber effects by nulling the angular clustering using random catalogs of the same (RA, Dec) as the observed galaxies.
Ref.\cite{bianchi/etal:2017,bianchi/verde:2020,bianchi/etal:2018,smith/etal:2019} up-weight targets by the inverse of the probability that a galaxy or a pair of galaxies is observed and apply the methods to DESI mock catalogs.
Ref.\cite{hahn/etal:2017} and \cite{yang/etal:2019} improve the nearest neighbor method by modeling the line-of-sight (LOS) distribution of fiber-collided objects and by considering the multiple fiber-collided objects, respectively.
We refer the reader to ref.\cite{bianchi/etal:2017} for comparison of the different methods.

In future/on-going deep fiber-fed spectroscopic surveys such as the PFS and DESI, the situation is more complicated than SDSS/BOSS for two reasons.
First, fibers can only move within the small patrol areas of robotic arms to place fibers in the PFS and DESI, while the fibers can be placed anywhere unless they do not collide with other fibers in SDSS/BOSS.
This leads to a non-random selection of galaxies and suppresses the clustering amplitude on all scales \cite{sunayama/etal:2020}.
We can achieve random sampling if we feed fewer parent photometric galaxies than targets, but in that case, the fiber allocation efficiency (the fraction of fibers which can find target), hence the number of observed galaxies, decreases.
Second, the wider redshift coverage of PFS ($0.6 < z < 2.4$) and DESI ($0.6 < z < 1.6$ for emission line galaxies) than SDSS/BOSS complicates the correction of fiber allocation effects, because the assumptions used in most of the correction methods are violated.
For example, the nearest neighbor method assumes that two galaxies close in angular space are physically close, but it is less likely in the PFS and DESI.  
The correction method using the angular clustering of the parent photometric galaxies is also less effective, because the observed angular clustering is almost homogeneous and the true clustering is obscured due to the wide redshift range of these surveys.

Ref.\cite{sunayama/etal:2020} examined the effects of fiber assignment on the galaxy clustering in configuration space (i.e., galaxy correlation function), focusing on the PFS cosmology survey.
In this paper, we extend our analysis to the galaxy clustering in Fourier space (i.e., galaxy power spectrum) and the cross correlation between galaxies and the weak-lensing field.
The PFS cosmology survey will perform the spectroscopic galaxy survey using the photometric galaxies of Subaru HSC. 
The PFS and HSC surveys provide not only the clustering measurements of [OII] emission-line galaxies (ELGs) but also its cross-correlation with the weak lensing field \cite{makiya/etal:2021}.
To maximize the scientific outcome of the PFS cosmology program in tandem with the HSC survey, it is crucial to investigate the effect of the fiber assignment on the galaxy-lensing cross power spectrum.
We show how the fiber assignment affects the measurement of the power spectrum and cross power spectrum, and how we can mitigate them.
We newly find that the fiber assignment creates artificial correlations between different redshifts and results in a significant suppression of the cross power spectrum. 

This paper is organized as follows. 
In section \ref{sec:survey_spec}, we describe the specifications of the PFS cosmology survey, including the construction of the mock catalog and the details of fiber assignment algorithm.
In section \ref{sec:results}, we show the effect of fiber assignment on the galaxy power spectrum and the galaxy-lensing cross power spectrum, and how we can mitigate them.
We summarize and conclude in section~\ref{sec:summary}.

\section{Specifications of the PFS cosmology survey}
\label{sec:survey_spec}
The Subaru PFS is a multiplexed fiber-fed optical and near-infrared spectrometer mounted at the prime focus of the Subaru Telescope\cite{pfs_tamura}, which is expected to be ready by the fall of 2023. 
The spectrograph covers a wide range of wavelengths from 380 to 1260 nm in a single exposure, which enables the PFS cosmology program to map the three-dimensional distribution of [OII] ELGs at $0.6 < z < 2.4$ over 1200 ${\rm deg}^2$ \cite{pfs_takada}.
The focal plane of the Subaru PFS is equipped with 2394 robotically reconfigurable fibers distributed over the 1.25~degree wide hexagonal field-of-view (FoV) as illustrated in figure~\ref{fig:pfs_cobra}. With this large FoV, the Subaru PFS can efficiently observe roughly 4 million ELGs within 100 nights.

In this section, we provide an overview of the PFS cosmology survey design and summarize details of the fiber assignment in the PFS cosmology as well as the galaxy and lensing mock catalogs used in this paper.
We omit some details of the instrumental design specific to the Subaru PFS, which is discussed in ref.\cite{sunayama/etal:2020}. 
While we use the fiber assignment for the Subaru PFS cosmology survey as an example, the nature of the problem is common to any fiber-fed spectroscopic galaxy surveys.

\subsection{Survey design: tiling and fiber assignment}
\begin{figure}[]
	\begin{center}
		\includegraphics[width=0.84\textwidth]{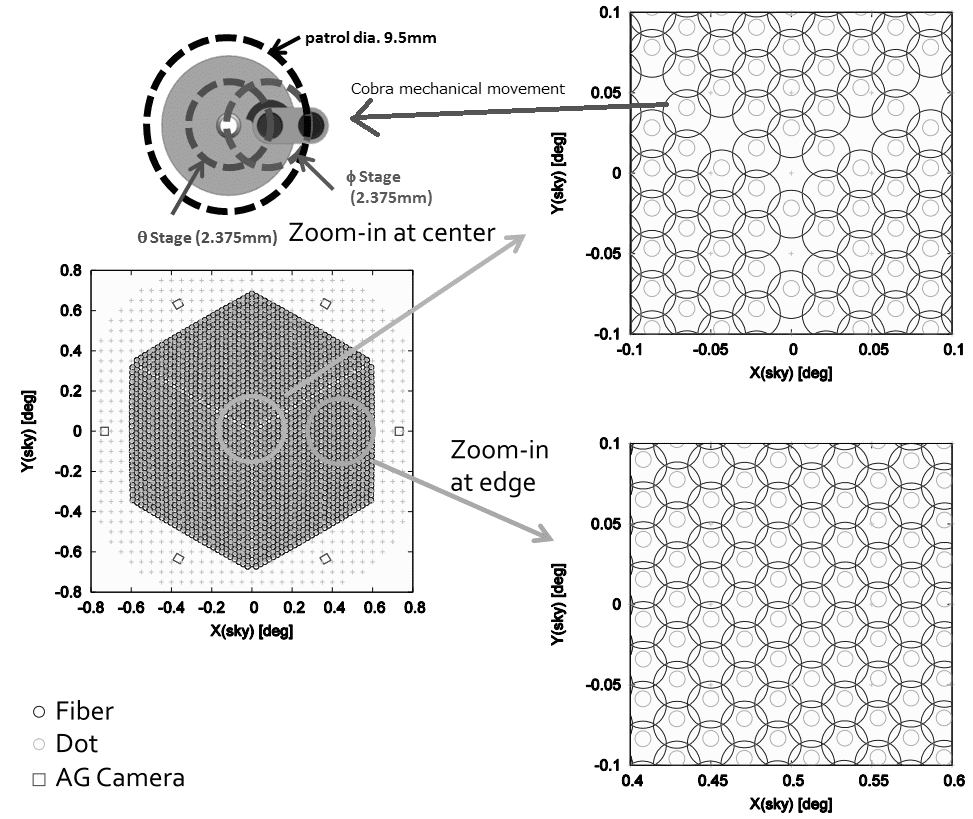}
		\includegraphics[width=0.4\textwidth]{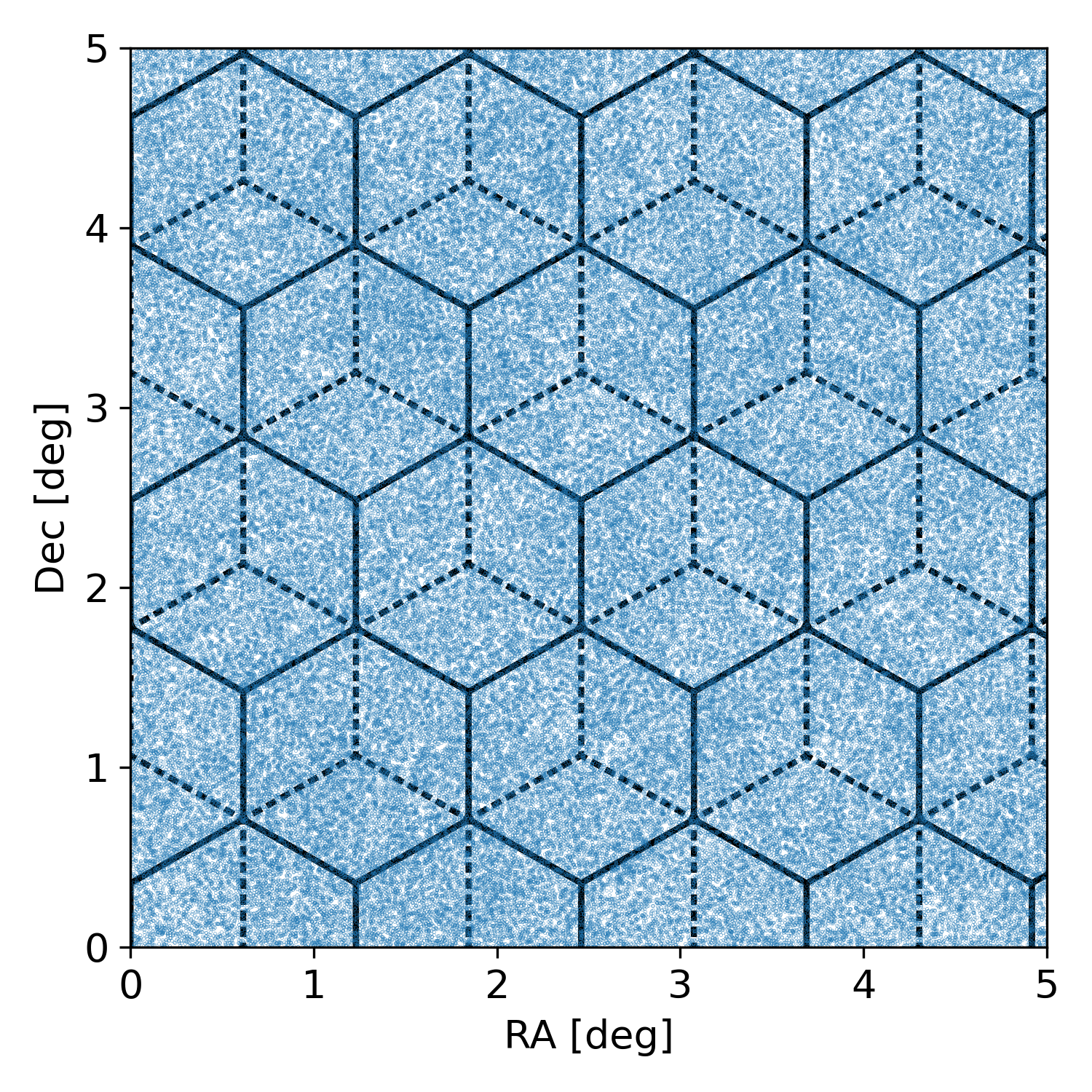}
		\caption{
		{\it Top}: An illustration of the configuration of fiber positioners on the focal plane of PFS (taken from ref.\cite{pfs_shimono}). The gray shaded hexagons in the left panel show the 1.25 ${\rm deg}^2$ FoV of the PFS.
		Each fiber is moved by two motors to cover its patrol area, as shown in the upper left panel. 
		Right panels show the configurations of the patrol area of fibers (black circles) at the center and the edge of FoV, respectively. 
		The gray circles show the ``dead'' region, where we can not observe targets.
		{\it Bottom}: The tiling strategy of PFS cosmology survey. Solid and dashed lines show the tiling of the first and second visits, respectively. The blue dots show the mock target galaxies at $0.6 < z < 2.4$.
		}
		\label{fig:pfs_cobra}
	\end{center}
\end{figure}

The PFS cosmology program aims to observe ELGs at $0.6 < z < 2.4$ within 100 nights covering the area of 1200 ${\rm deg}^2$ using a hexagonal plate. 
The bottom panel of figure~\ref{fig:pfs_cobra} illustrates the survey strategy: the hexagonal plates are uniformly placed and the center of the plate is dithered by the radius of the hexagon between the first and second visits. 
Throughout this paper, we call each hexagonal region in the bottom panel of figure~\ref{fig:pfs_cobra} a ``tile''.

To achieve the survey with a reasonably high number density of galaxies over the wide redshift range, maximizing the efficiency of fiber usage is crucial. 
However, it results in a low completeness, which is the ratio of the number of observed galaxies\footnote{In this paper we assume 100\% success rate of redshift determination for simplicity, i.e., the number of observed galaxies is equal to the number of fiber-assigned galaxies.} to the number of targets. 
In the nominal two-visit survey of PFS cosmology, we prepare roughly 8000 targets per field, i.e. $8000/2394 \simeq 3.3$ targets per fiber, to achieve the 90\% of fiber usage efficiency \cite{pfs_takada} and 54\% completeness. 
Having such low completeness suppresses the amplitude of galaxy clustering due to the loss of long-wavelength fluctuations, whose wavelength is larger than the tile size. 
This is because the fluctuations in the number of target galaxies per tile are lost by a fixed number of fibers. 
This was not the problem for the previous spectroscopic surveys such as BOSS because the number of target galaxies was less than the number of available fibers. 
The details of the problem and the mitigation scheme are discussed in section 4.1 of ref.\cite{sunayama/etal:2020}.

Another difference between the PFS and the previous spectroscopic surveys is that each fiber positioner of PFS has a finite “patrol” area within which the fiber can move to the desired target as shown in the top panel of figure~\ref{fig:pfs_cobra}. 
A finite patrol area together with low completeness results in a non-uniform sampling of galaxies. 
When a positioner has more than two targets within its patrol area, only two galaxies are observed in the nominal two-visit survey. 
This causes a non-random sampling of target galaxies, which further suppresses the clustering amplitude on small scales.

\subsection{Light-cone mock catalogs for galaxies and lensing map}
To investigate the fiber assignment artifacts on the galaxy power spectrum and the galaxy-lensing cross power spectrum, we generate the mock galaxy catalog and weak lensing map by using the public code {\tt lognormal\_lens} \cite{makiya/etal:2021}\footnote{\url{https://wwwmpa.mpa-garching.mpg.de/~komatsu/codes.html}.}.

The code first generates a series of three-dimensional matter density fields on a regular grid from the input power spectra assuming that density fields follow a log-normal probability density function.
To obtain the density fluctuation field $\delta({\bm x})$, the code first generates a log-transformed density field $\ln{[1+\delta({\bm x})]}$ which is a Gaussian random field characterized by $\xi^{G}$ \cite{coles/etal:1991},
\begin{equation}
\xi^{G}(r) = \ln{[1+\xi(r)]},
\end{equation}
where $\xi(r)$ is the input correlation function.
The code then generates $\delta({\bm x})$ by exponentiating the log-transformed density field.

Next, the code constructs a single light-cone from a series of three-dimensional matter density fields and ray-traces a light-cone to obtain the weak lensing convergence field.
Ray-tracing is done by the public code RAYTRIX \cite{raytrix}. 
First, we project a three-dimensional density field $\delta$ into a two-dimensional field using the plane-parallel approximation, 
\begin{equation}
    \label{eq:projection}
	\delta^{\rm proj}(x,y) = \frac{1}{N_{{\rm grid},z}}\sum_{z_i}\delta(x,y,z_i),
\end{equation}
where $N_{{\rm grid},z}$ is the grid number along the LOS.
We shall call this projected density field a ``mass sheet''.
Next, the two-dimensional deflection potential $\Psi$ is computed from the mass sheet as
\begin{equation}
\label{eq:poisson}
	\nabla^2 \Psi(x,y) = \frac{3\Omega_m H_0^2}{c^2} \delta^{\rm proj}(x,y),
\end{equation}
where $\Omega_m$ is the matter density parameter, $H_0$ is the Hubble parameter and $c$ is the speed of light.
From the deflection potential, a Jacobian of lensed to unlensed coordinates, $A_n$, is computed as \begin{equation}
    A_n = I-\sum_{i=1}^{n-1}
    \frac{\chi_i(\chi_n-\chi_i)}{a(\chi_i)\chi_n}
    U_i A_i,
\end{equation}
and
\begin{equation}
    U_i =      \begin{pmatrix}
    \displaystyle \frac{\partial^2 \Psi^i}{\partial x_1 \partial x_1} & 
    \displaystyle \frac{\partial^2 \Psi^i}{\partial x_1 \partial x_2}\\
    &\\
    \displaystyle \frac{\partial^2 \Psi^i}{\partial x_2 \partial x_1} & 
    \displaystyle \frac{\partial^2 \Psi^i}{\partial x_2 \partial x_2}\\
    \end{pmatrix},
\end{equation}
where $I$ is the identity matrix, $\Psi^i$ is the deflection potential of the $i$-th mass sheet, $n$ is the total number of mass sheets, $a$ is the scale factor, and $\chi_n$ and $\chi_i$ are the comoving distances to the source plane and the $i$-th mass sheet, respectively.
The source plane is located at the $n$-th mass sheet.
The Jacobian $A_n$ is related to the lensing convergence $\kappa$ and shear field $\gamma  = (\gamma_1,\gamma_2)$ as
\begin{equation}
    A_s = 
    \begin{pmatrix}
    1-\kappa-\gamma_1 && -\gamma_2\\
    -\gamma_2 && 1-\kappa+\gamma_1\\
    \end{pmatrix}.
\end{equation}
The convergence $\kappa$ and shear $\gamma$ characterize magnification and stretching of the source image around the lens, respectively.
In the following, we focus on $\kappa$.

The code also generates galaxies by Poisson sampling the underlying galaxy density field, which is also generated from the input galaxy power spectrum assuming that the galaxy density field follows the log-normal distribution.
We use the same initial random seed for matter and galaxy density fields, to ensure that the matter and galaxy fields correlate.
We also generate the peculiar velocity field of galaxies from the underlying matter density field by using the linear continuity equation.
The galaxy density field is generated on the regular grid. We randomly populate galaxies within each grid cell, while assigning the same velocity to galaxies within the same grid.
We refer the readers to ref.\cite{makiya/etal:2021} and ref.\cite{agrawal/etal:2017} for further details of our lognormal simulations.

The input power spectra are generated by the public code {\tt CLASS} \cite{class1,class2}, assuming a flat $\Lambda$CDM model with the parameters of {\it Planck} 2015 `TT,TE,EE+lowP': $\Omega_b h^2 = 0.02225$, $\Omega_c h^2 = 0.1198$, $n_s = 0.9645$, $\ln(10^{10} A_s) = 3.094$ and $h = 0.67021$ with the minimum neutrino mass of $M_\nu = 0.06$ [eV] \cite{planck2015}. 
For simplicity, we use the linear matter power spectrum and assume the linear galaxy bias.
Following ref.\cite{pfs_takada}, we assume that the linear galaxy bias is given by $b(z) = 0.9+0.4z$.
The source galaxies of the weak lensing convergence field are assumed to be located at $z = 3.0$.
We ignore the shape noise in the lensing measurement because it is not affected by the fiber assignment.

\begin{figure}[]
	\begin{center}
		\includegraphics[width=1.0\textwidth]{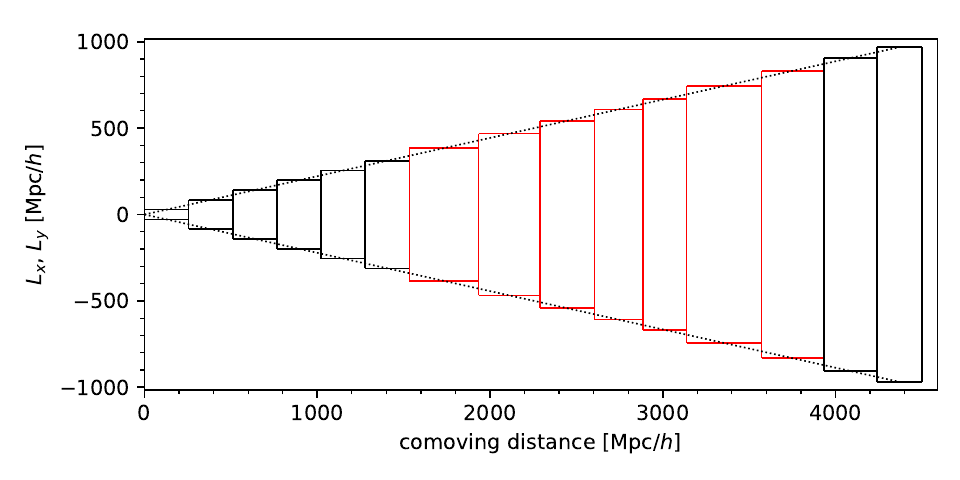}
		\caption{Configuration of the mock light cone.
		The matter density field is generated in each solid box. 
		The red boxes also contain galaxies mocking the spectroscopic sample of PFS cosmology survey.
		The black dotted lines show the $25\times25$ ${\rm deg}^2$ FoV.
		The side lengths of each box are summarized in table~\ref{tb:survey}.}
		\label{fig:light_cone}
	\end{center}
\end{figure}

The configuration of matter density fields and light-cone is shown in figure~\ref{fig:light_cone}.
The FoV of the simulation (shown in the black dotted lines in the figure \ref{fig:light_cone}) is $25 \times 25 = 625 \;{\rm deg}^2$, similar to the ``Fall equatorial field'' of the PFS cosmology, $630\;{\rm deg^2}$.
The configuration of three-dimensional boxes is determined as follows.
First we prepare boxes in which the PFS spectroscopic galaxy samples are generated (the red boxes in figure \ref{fig:light_cone}) with the redshift interval of $\Delta z$ = 0.2 for $0.6 < z < 1.6$ and  $\Delta z = 0.4$ for $1.6 < z < 2.4$, following the fiducial PFS survey parameters given in ref.\cite{pfs_takada}.
Next, we prepare other foreground and background boxes (the black boxes in figure \ref{fig:light_cone}).
The number of boxes is determined so that each box has a similar length along the LOS, $L_z$. 
The maximum redshift $z_{\rm max} = 3.2$ is chosen to match the assumed redshift distribution of source galaxies of the HSC survey.
The box side-lengths perpendicular to the LOS, $L_x$ and $L_y$, are chosen to match the opening angle of the simulation times comoving distances (the horizontal axis) to the box centers.

Each lognormal box is generated with the three-dimensional grid-number of $N_{\rm grid,3D} = 256$ for the axis perpendicular to LOS, while the grid number along the LOS is determined so that each grid has a cubic shape.
The number of two-dimensional angular grids for the lensing map is also set to $N_{\rm grid,2D} = 256$, i.e., the grid size is $25.0/256 = 0.098$ degrees on a side.

The box geometry, galaxy number density, and the linear bias parameter of our simulation are summarized in figure~\ref{fig:light_cone} and table~\ref{tb:survey}.

\begin{table}[hbt]
\begin{center}
  \begin{tabular}{cccccc}
    redshift & $L_{x,y}$ & $L_z$ & $\bar{n}_g^{\rm true}$ & $\bar{n}_g^{\rm obs}$ & $b_g$ \\
    		 & [Mpc/$h$] & [Mpc/$h$] & [$10^{-4}h^3 {\rm Mpc^{-3}}$] & [$10^{-4}h^3 {\rm Mpc^{-3}}$]      \\
	\hline
	$0.0 < z < 0.09$ & 56.7 & 255.6 & -- & -- & -- \\
	$0.09 < z < 0.18$ & 170.0 & 255.6 & -- & -- & -- \\
	$0.18 < z < 0.27$ & 283.3 & 255.6 & -- & -- & -- \\
	$0.27 < z < 0.38$ & 396.6 & 255.6 & -- & -- & -- \\ 
	$0.38 < z < 0.48$ & 510.0 & 255.6 & -- & -- & -- \\
	$0.48 < z < 0.6$ & 623.3 & 255.6 & -- & -- & -- \\ 
    $0.6 < z < 0.8$ & 768.8 & 400.6 & 5.4 & 2.7 & 1.18 \\
    $0.8 < z < 1.0$ & 937.2 & 354.7 & 17.2 & 8.4 & 1.26 \\
    $1.0 < z < 1.2$ & 1084.8 & 315.2 & 16.6 & 8.2 & 1.34 \\
    $1.2 < z < 1.4$ & 1217.0 & 281.4 & 22.4 & 10.9 & 1.42 \\
    $1.4 < z < 1.6$ & 1335.4 & 252.5 & 15.8 & 7.7 & 1.5 \\
    $1.6 < z < 2.0$ & 1487.7 & 434.5 & 7.2 & 3.5 & 1.62 \\
    $2.0 < z < 2.4$ & 1664.0 & 360.8 & 7.8 & 3.8 & 1.78 \\
	$2.4 < z < 2.8$ & 1811.6 & 305.1 & -- & -- & -- \\ 
	$2.8 < z < 3.2$ & 1937.3 & 262.0 & -- & -- & -- \\ 
    \hline
  \end{tabular}
  \caption{The details of the simulation boxes generated to construct the mock light-cone.
  $L_{x,y}$ is the comoving side lengths of the box perpendicular to the LOS direction, while $L_z$ is the length along the LOS.
  $\bar{n}_g^{\rm true}$ and $\bar{n}_g^{\rm obs}$ denote the mean number density of target galaxies before and after the fiber assignment, respectively.
  $b_g$ is the linear galaxy bias.}
  \label{tb:survey}
  \end{center}
\end{table}

\subsection{Exposure Targeting Software (ETS)}
To simulate the fiber assignment scheme in the PFS cosmology survey, we use the Exposure Targeting Software (ETS)\footnote{\url{https://github.com/Subaru-PFS/ets_fiberalloc}.} which is developed for the PFS \cite{pfs_shimono}.

As illustrated in figure~\ref{fig:pfs_cobra}, PFS has 2394 fibers within a hexagonal FoV of $\sim 1.25 {\rm deg}^2$.
The fiber positioner, ``Cobra'', consists of two rotational stages to cover a circular patrol area per fiber.
These positions on the focal plane are fixed.
The ETS takes angular positions (RA, Dec) and priorities of target galaxies as inputs, and finds targets in the patrol area of each fiber.
We adopt the plane-parallel approximation for each simulation box when projecting the three-dimensional galaxy distribution to the (RA, Dec) space.
If there are multiple galaxies in the same patrol area, the ETS selects a galaxy based on the given priority and the position of galaxies within the patrol area.
In this paper, we employ a ``naive'' algorithm, in which the fiber is simply assigned to a target galaxy with the highest priority.
If there are multiple galaxies with the highest priority, the algorithm randomly selects one of them.
Here we assign the same priority to all target galaxies, which is appropriate for the PFS cosmology program.
We could also assign the priority according to physical properties of target galaxies such as the expected line flux, for example, in order to increase the success rate. 
However, in that case the fiber assignment effects would correlate with the physical properties of target galaxies and would become more complicated.
As shown in the bottom panel of figure~\ref{fig:pfs_cobra}, we place the hexagonal tiles uniformly, while separating the tile centers for the first and second visit by the radius of the hexagonal tile.

We generate 500 realizations of {\tt lognormal\_lens} simulation and run the ETS on each mock.
In the following, we call the galaxy catalog without fiber assignment the ``true'' sample, while those with fiber assignment the ``observed'' sample. 
The mean number density of galaxies in the observed sample is given in table~\ref{tb:survey}.                        

\section{Fiber assignment artifacts}
\label{sec:results}
In this section, we present the impact of fiber assignment on the galaxy power spectrum and the galaxy-lensing cross power spectrum as well as the mitigation scheme.

We show that the observed galaxy power spectrum is suppressed at all scales in the same manner as for the correlation function in configuration space \cite{sunayama/etal:2020}.
We will also show that the suppression is localized at the $\mu_k = 0$ modes, where $\mu_k$ is the cosine between the wave vector and the LOS direction.

Next, we show that the observed galaxy-lensing cross power spectrum is also suppressed and even becomes negative at high redshifts.
This is due to the artificial correlation between galaxies at different redshift slices introduced by fiber assignment.

To mitigate the fiber assignment artifacts, we apply the so-called ``Individual-Inverse Probability'' (IIP) method \cite{bianchi/etal:2017} in which observed galaxies are up-weighted by the inverse probability to be observed.
We find that the IIP method recovers the true power spectrum at accuracy better than 1\% up to $k = 0.2$ [$h$/Mpc].
If we further take into account the pairwise probability of galaxies to be observed, we recover the true spectrum at $< 1\%$ accuracy up to the Nyquist frequency.
On the other hand, the IIP method recovers the galaxy-lensing cross power spectrum at an accuracy better than 1\% up to the Nyquist frequency.
The IIP method is sufficient for the cross power spectrum since the pairwise probability does not enter in this statistics.

\subsection{Galaxy power spectrum in real space}
\label{subsec:power}
First, we investigate the galaxy power spectrum in real space (i.e., ignoring the peculiar velocity of galaxies).
In the following, all of the power spectrum and cross power spectrum measurements are done by using the Fast Fourier transform (FFT).
We use the FFTW library for Fourier transform \cite{FFTW05}.
The nearest grid point (NGP) assignment is adopted to compute the galaxy number density contrast\footnote{In general, the NGP density assignment scheme is not reliable already at the scales larger than the grid size. However, as investigated in Ref.\cite{agrawal/etal:2017,makiya/etal:2021}, the NGP assignment recovers the input power spectrum up to nearly the Nyquist frequency in the lognormal simulation as long as we use the same mesh size as the realization of lognormal field.}, $\delta_g({\bm x}) = n_g({\bm x})/\bar{n}_g - 1$, where $n_g({\bm x})$ is the galaxy number density estimated within the cubic grid at a three-dimensional position ${\bm x}$.
The grid size is the same with lognormal realizations of galaxies.
The mean galaxy number density $\bar{n}_g$ is defined as $\bar{n}_g = N_g/V$, where $N_g$ is the total number of galaxies within survey volume $V$.
As we 

We estimate the power spectrum as
\begin{equation}
	P(k,\mu_k) = \frac{1}{N_{\rm k}}\sum_{i=1}^{N_k} 
	\left[ \tilde{\delta}_g({\bm k_i})\tilde{\delta}_g^{*}({\bm k_i})-P_{\rm shot} \right],
\end{equation}
where $\tilde{\delta}_g({\bm k})$ is the Fourier transformed galaxy number density contrast and $N_{k}$ is the number of Fourier modes within a given $(k,\mu_k)$ bin.
The cosine of the angle between ${\bm k}$ and the LOS ($z$-direction), $\mu_k$, is defined as $\mu_k \equiv k_z/k$.
Here we assume that the LOS vector is the same for all galaxies (i.e., global plane-parallel approximation).
The shot noise power spectrum $P_{\rm shot}$ is given by $P_{\rm shot} = 1/\bar{n}_g$.
Since the galaxy density fields are generated to satisfy the periodic boundary condition, we do not need to care about the window effect of the survey boundary.

\begin{figure}[]
	\begin{center}
		\includegraphics[width=1.0\textwidth]{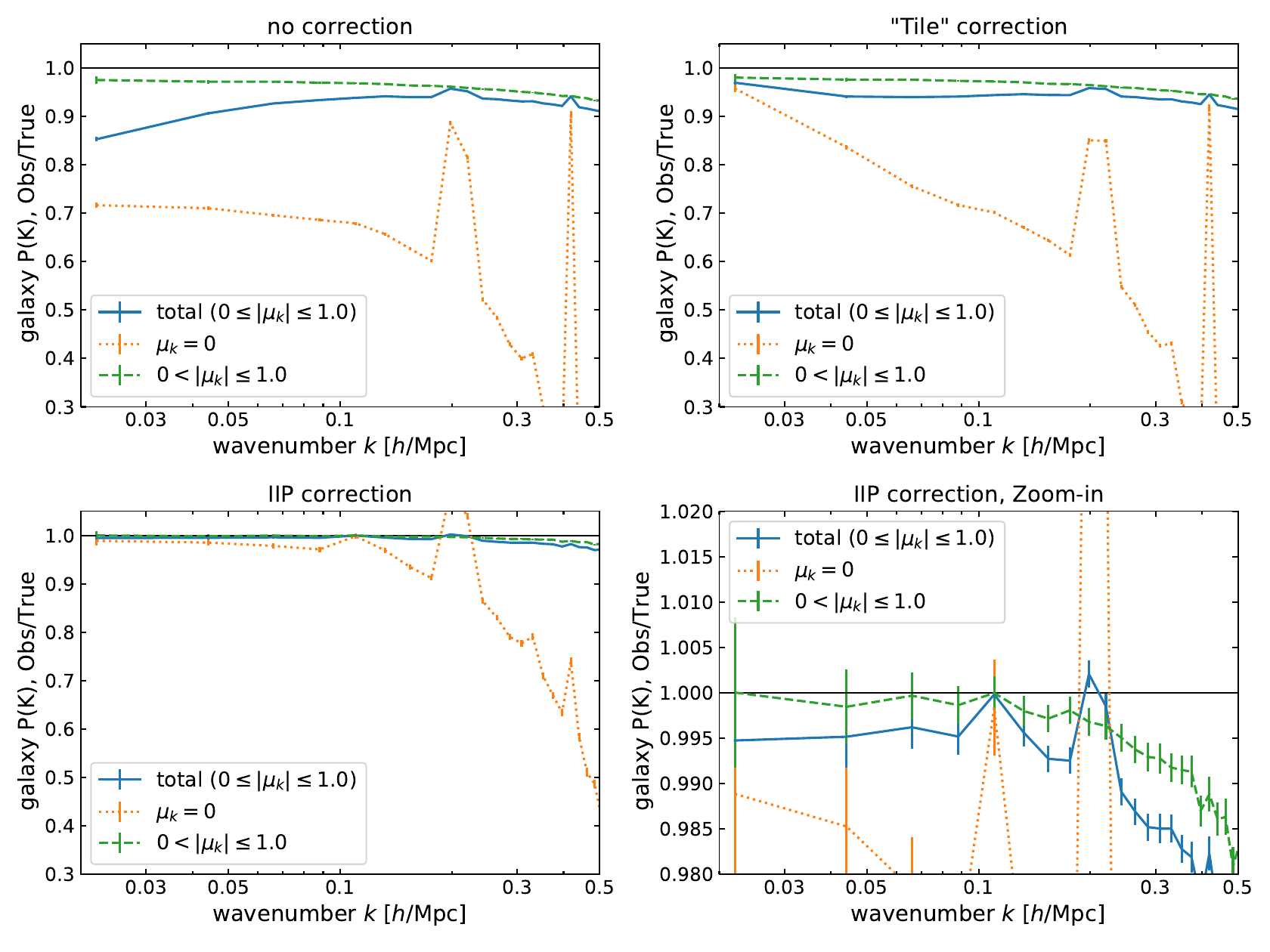}
		\caption{(Top-Left) The ratio of the ``observed'' to ``true'' galaxy power spectra.
		The blue solid line is the total power spectrum, while the orange dotted line and the green dashed line are the spectra with $\mu_k = 0$ and $0.0 <|\mu_k | \leq 1.0$, respectively.
		The data points are the mean of 500 realizations.
		The error bars are the standard deviation of mean, i.e. the standard deviation divided by the square root of the number of realizations.
	    (Top-Right) The same as Top-Left, but the tile weighting is applied. 
		(Bottom-Left) The same as Top-Left, but the IIP weighting is applied.
		(Bottom-Right) Zoom-in view of Bottom-Left panel.
		}
		\label{fig:pkmu_real}
	\end{center}
\end{figure}

The top-left panel of figure~\ref{fig:pkmu_real} shows the ratio of the ``observed'' to ``true'' galaxy power spectra at $z = 1.3$.
Here we show the angle averaged power spectrum as well as the power spectra at $\mu_k = 0$ and $0 < \mu_k \leq 1$.
The wave vector $k$ is binned in the interval of the fundamental frequency, $\Delta k = 0.022$ [$h$/Mpc].
Figure~\ref{fig:pkmu_real} clearly shows that the most of the fiber assignment effect is localized at $\mu_k = 0$.
A similar trend is also seen in the fiber-assigned power spectrum of the DESI-like survey \cite{pinol/etal:2017}.
This result can be understood as follows.
The $k_z = 0$ ($\mu_k = 0$) component of the Fourier-transformed density field $\tilde{\delta}$ is expressed as
\begin{eqnarray}
    \tilde{\delta}(k_x,k_y,k_z=0) &=& \int\int\int
    \delta(x,y,z)e^{-i(k_x x+k_y y)} \;{\rm d}x\;{\rm d}y\;{\rm d}z \nonumber \\
    &=& \int\int \left[\int  \delta(x,y,z) \;{\rm d}z\right]
    e^{-i(k_x x+k_y y)}{\rm d}x\;{\rm d}y.
\end{eqnarray}
This equation shows that the $\mu_k = 0$ mode of $\delta(k,\mu_k)$ corresponds to the projected density field along the LOS. 
Since the fiber assignment is based on the angular positions of galaxies, the mode corresponding to the projected density field is affected the most.

In ref.\cite{sunayama/etal:2020} we found that the suppression at large scales is caused by the ``tiling effect''.
Since the number of fibers within each tile is fixed, the correlation amplitude at a scale larger than a tile is suppressed when the number of target galaxies is much larger than the number of fibers.
To correct for this tiling effect, ref.~\cite{sunayama/etal:2020} used the weight given by
\begin{equation}
\label{eq:tile}
    w_i^{\rm tile} = \frac{N_{{\rm target},i} }{N_{{\rm obs},i}},
\end{equation}
where $N_{{\rm target},i}$ and $N_{{\rm obs},i}$ are the number of target and observed galaxies within each tile, respectively.
The top-right panel of figure~\ref{fig:pkmu_real} shows the power spectra after applying the weights.
We can see that the weight recovers the true power spectrum at large scales. 

The suppression at small scales is caused by the under-sampling of over-dense regions due to the fiber assignment, which is also seen in ref.\cite{sunayama/etal:2020}.
Since the fibers are uniformly distributed within the FoV but the target galaxies are not, we miss a certain fraction of targets in the dense region while we observe most of the targets in the under-dense region.
The PFS cosmology program is designed to have a high number density of target galaxies to achieve a high fiber usage efficiency.
This results in a high probability of having many galaxies in a single patrol area, hence the significant suppression of the observed power spectrum.

To correct for the fiber assignment effects, we apply the IIP method following ref.\cite{sunayama/etal:2020}. 
First, we run the ETS 100 times for each realization of the mock and estimate the probability of the $i$-th galaxy to be observed, $p_{i}$, for all galaxies.
Next we up-weight each observed galaxy by $w_{i} = 1/p_{i}$.
In this case, the galaxy number density contrast is computed as $\delta_g({\bm x}) = n_{g,{\rm IIP}}({\bm x})/\bar{n}_{g,{\rm IIP}}-1$, where $n_{g,{\rm IIP}}({\bm x})$ is the local mean of $w_{i}$ within the grid at ${\bm x}$, and $\bar{n}_{g,{\rm IIP}} = \sum_{i=1}^{N_g} w_{i}/V$ is the global mean of $w_{i}$. 
Note that the IIP method also corrects the tiling effect, since the tile-to-tile variation of the number of target galaxies is naturally included in the variation of $w_{\rm IIP}$.
Following ref.\cite{fkp,yamamoto/etal:2006,bianchi/verde:2020}, the shot noise of weighted galaxies is computed as
\begin{equation}
P_{\rm shot, IIP} = \frac{\sum_{i=1}^{N_g} w_i^2}{\bar{n}_{g,{\rm IIP}}^2 V} 
= \frac{\sigma[w]^2+\bar{w}^2}{\bar{w}^2}P_{\rm shot},
\label{eq:shot}
\end{equation}
where $\bar{w}$ and $\sigma[w]$ are defined as $\bar{w} = \sum_{i} w_i/N_g$ and $\sigma[w]^2 = \sum_i w_{i}^2/N_g-\bar{w}_i^2$, respectively.
This equation shows that the shot noise is enhanced in the case of non-uniform sampling, i.e. $\sigma[w_{\rm IIP}] > 0$.
The bottom panels of figure~\ref{fig:pkmu_real} show the ratio of the observed to true galaxy power spectra for the IIP weighting. 
As in the case of the galaxy correlation function discussed in ref.\cite{sunayama/etal:2020}, the IIP weighting scheme recovers the true power spectrum at an accuracy better than $1\%$ up to $k = 0.2$ [$h$/Mpc].

The spike-like features seen in the observed power spectra are caused by the way we place the hexagonal tiles.
Since the fibers cannot be placed at the edge of the tile, there are small gaps between tiles where galaxies are not observed at all and these gaps generate artificial clustering feature.
As shown in figure~\ref{fig:pfs_cobra}, the angular separation of the gaps is roughly the same with the angular radius of tile ($\sim0.7$ deg), which corresponds to the comoving length of $\sim34.0$ [Mpc/$h$] and the wave number of $k \sim 0.18$ [$h$/Mpc]) at $z=1.3$.
The positions of the spike, $k\sim0.2$ and $\sim0.4$ [$h$/Mpc], roughly match that wavelength and their integer multiples, respectively.
The effect of such a non-uniform survey window can be mitigated by using, e.g, a random catalog \cite[e.g.,][]{beutler/etal:2014}.

\begin{figure}[]
	\begin{center}
		\includegraphics[width=0.49\textwidth]{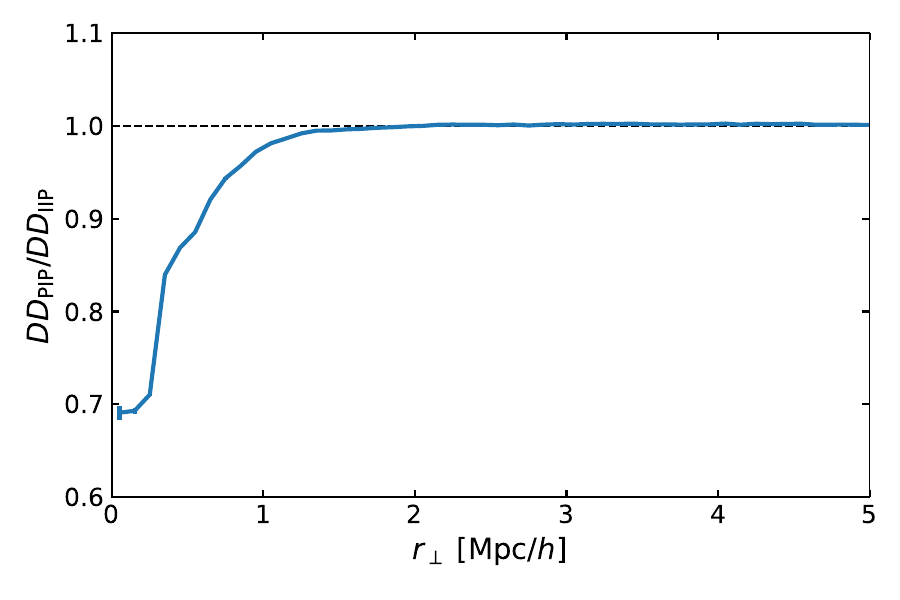}
        \includegraphics[width=0.49\textwidth]{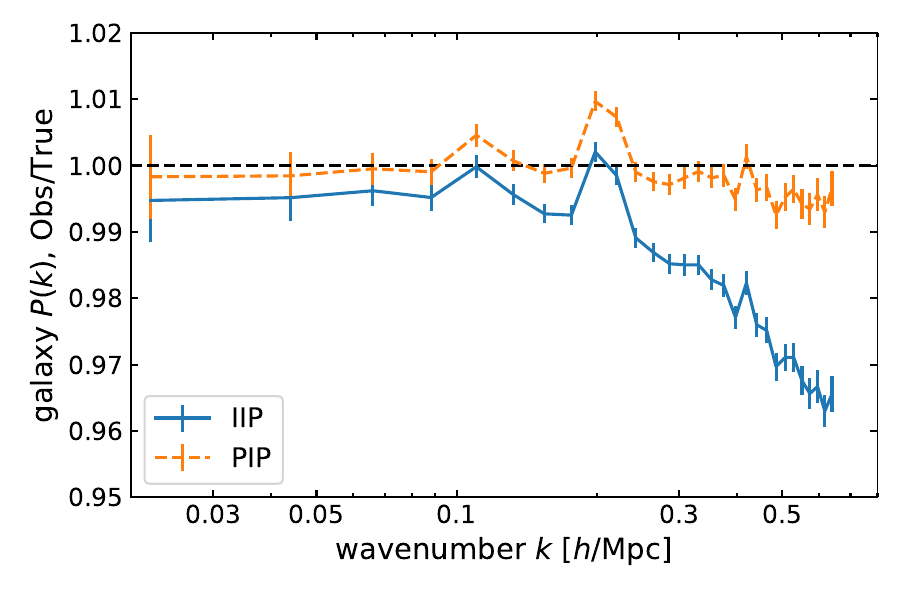}
		\caption{(Left) The ratio of the galaxy pair count corrected by the PIP weights, to that corrected by the IIP weights, as a function of the separation perpendicular to the LOS, $r_{\perp}$, at $z = 1.3$. 
		The data points show the mean and standard deviation of 10 realizations.
	    (Right) The ratio of the ``observed'' to ``true'' power spectra at $z = 1.3$, corrected by the IIP and PIP weights, respectively. 
	    The data points show the mean and standard deviation of 100 realizations.
		}
		\label{fig:pip}
	\end{center}
\end{figure}

The IIP correction does not fully recover the ``true'' power spectrum at small scales.
This is because the IIP estimator is not unbiased at a small scale where the correlation between $p_{i}$ is not negligible.
Several studies \cite[e.g.][]{bianchi/etal:2017,bianchi/verde:2020} show that the so-called ``Pairwise-Inverse Probability'' (PIP) method, in which galaxy pairs are up-weighted by the pairwise probability to be observed, can unbias the galaxy spectrum multipoles at smaller scales than the IIP method.
In the left panel of figure~\ref{fig:pip} we show the ratio of the PIP corrected to the IIP corrected galaxy pair count, as a function of the separation perpendicular to the LOS, $r_{\perp}$.
It is clearly shown in the figure that the ratio deviates from unity at $r_{\perp} \lesssim 2.0\;[{\rm Mpc}/h]$.
Following ref.\cite{bianchi/etal:2017,bianchi/verde:2020}, we can express the PIP corrected power spectrum as
\begin{eqnarray}
    \label{eq:pip}
    P^{\rm PIP}(k) &=& \frac{1}{I}\int \frac{d\Omega_k}{4\pi}
    \left[\sum_{ij}(w_{ij}-w_{i}w_{j})e^{i{\bm k}({\bm x}_i-{\bm x}_j)}\right]
    +P^{\rm IIP}(k) \nonumber \\
    &=& \frac{1}{I}
    \sum_{ij}(w_{ij}-w_{i}w_{j})j_0(k|{\bm x}_i-{\bm x}_j|)
    +P^{\rm IIP}(k),
\end{eqnarray}
where $I = \bar{n}^2_{g,{\rm IIP}}V$ is the normalization constant, $d\Omega_k$ is the solid angle element in $k$-space, and $w_{ij} = 1/p_{ij}$ is the pairwise weight for a pair of the $i$-th and $j$-th galaxies.
In the second raw of the equation, the angular integration in $k$-space is reduced to the spherical Bessel functions of the first kind, $j_\ell$.
The right panel of figure~\ref{fig:pip} shows the power spectrum corrected by the PIP method. 
To reduce the computational cost, we assume that $w_{ij} -w_i w_j = 0$ at $r_{\perp} > 3.0\;[{\rm Mpc}/h]$.
Figure~\ref{fig:pip} clearly shows that the PIP method recovers the true power spectrum at an accuracy better than $1\%$ at all scales.

\begin{figure}[]
	\begin{center}
		\includegraphics[width=1.0\textwidth]{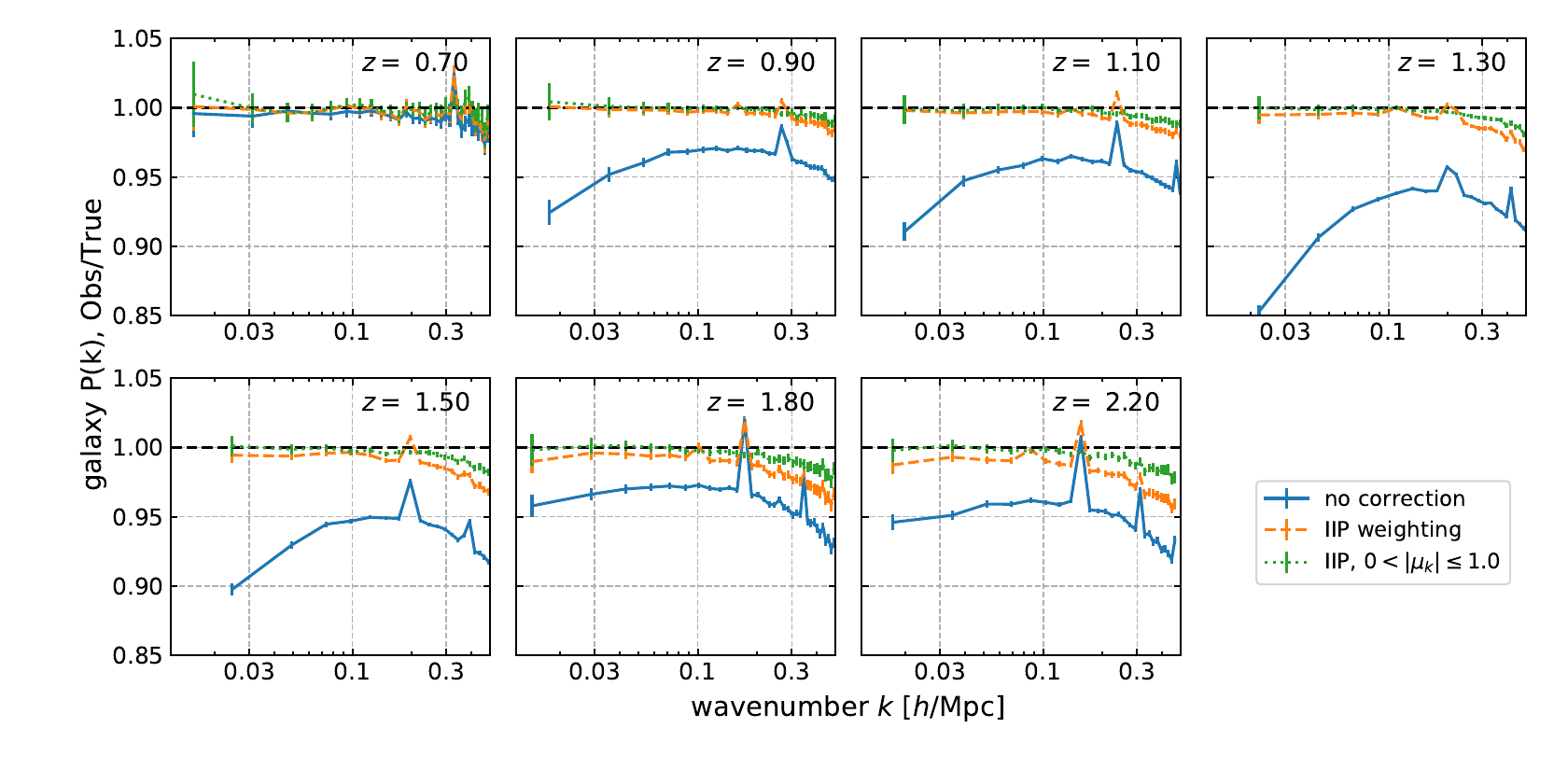}
    		\caption{The ratio of the ``observed'' to ``true'' galaxy power spectra for all redshift bins of the PFS survey. 
    		The blue solid lines show the results without any correction, while the orange dashed lines show the results corrected by the IIP weights. 
    		The green dotted lines show the corrected by the IIP weights and the $\mu_k = 0$ mode is removed.}
		\label{fig:pk_gg_real}
	\end{center}
\end{figure}

Figure \ref{fig:pk_gg_real} shows the ratio of the observed to true galaxy power spectra in real space for all redshift bins, as well as those corrected by the IIP weighting.
We also show the results when we remove the $\mu_k = 0$ mode, which is a major cause of suppression at small scales.
We find that the IIP weighting recovers the true power spectra at $\sim 1\%$ accuracy up to $k = 0.2$ [$h$/Mpc] for all redshifts.
This accuracy improves to better than $1\%$ if we remove the $\mu_k = 0$ mode.
The figure shows that the fiber effect is not strong at the lowest redshift bin, $0.6 < z < 0.8$.
This is due to its low number density of target galaxies.
When the small number of galaxies is considered as targets, most of the fibers have less than two targets in the patrol area.
In that case, the selection of galaxies becomes closer to random and therefore the fiber assignment has little effect.

\subsection{Galaxy power spectrum in redshift space}
Here we investigate the galaxy auto power spectrum in redshift space.
First we displace the observed galaxies by the LOS component of their peculiar velocity using the plane-parallel approximation, and then compute the multipole expansion of the power spectrum $P_{\ell}(k)$ defined as 
\begin{equation}
    P(k,\mu_k) = \sum_{\ell = 0}^{\infty} P_{\ell}(k)\mathcal{L}_{\ell}(\mu_k),
\end{equation}
where $\mathcal{L}_{\ell}$ is the $\ell$-th order Legendre polynomial.

\begin{figure}[]
	\begin{center}
		\includegraphics[width=1.0\textwidth]{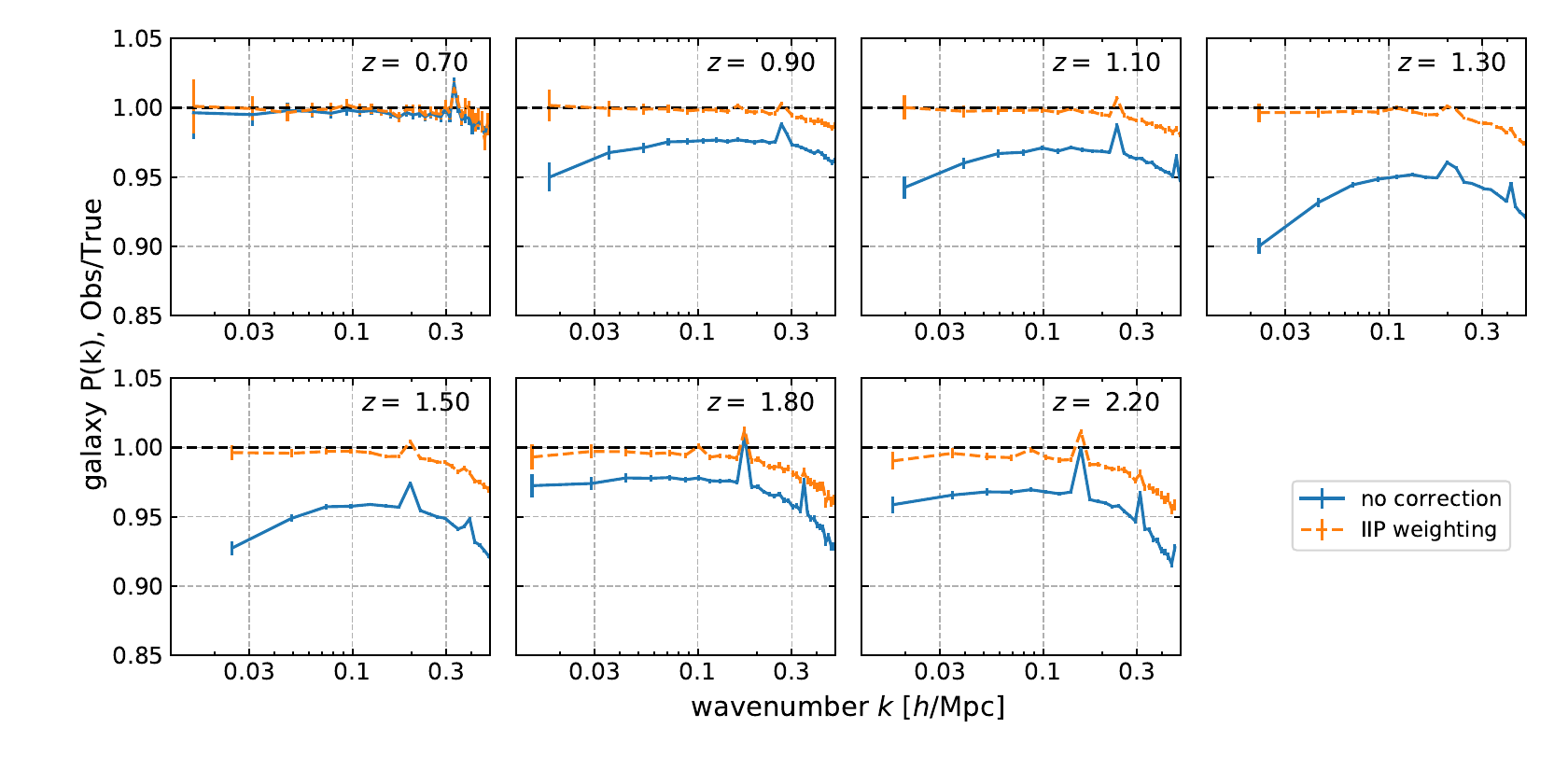}
		\caption{
		The same as figure~\ref{fig:pk_gg_real}, but for the monopole power spectra in redshift space.}
		\label{fig:pk_gg_mono}
	\end{center}
\end{figure}

\begin{figure}[]
	\begin{center}
		\includegraphics[width=1.0\textwidth]{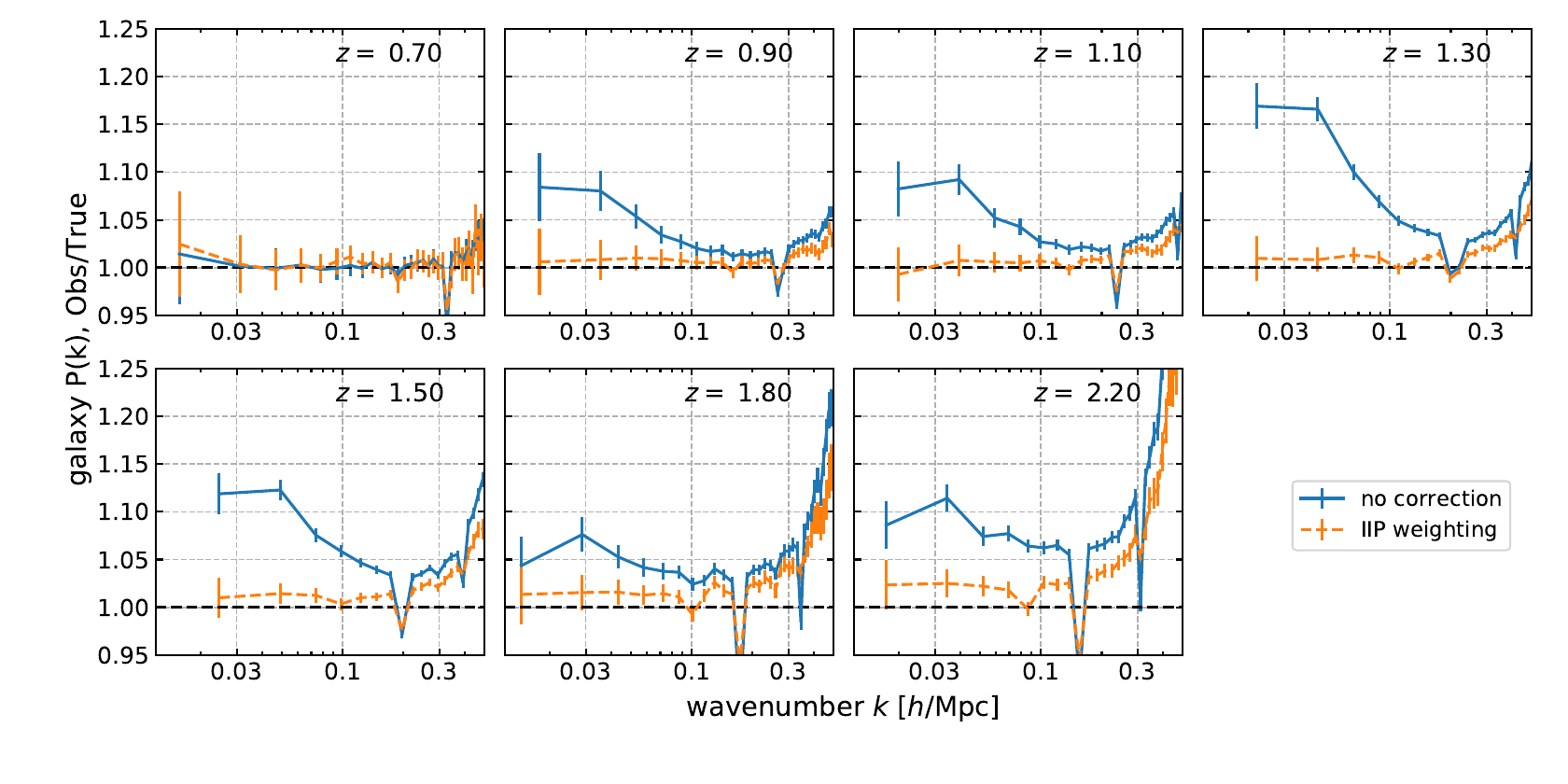}
		\caption{The same as figure~\ref{fig:pk_gg_real}, but for the quadrupole power spectra in redshift space.}
		\label{fig:pk_gg_quad}
	\end{center}
\end{figure}

Figure~\ref{fig:pk_gg_mono}~and~\ref{fig:pk_gg_quad} show the ratio of the observed to true galaxy power spectra for monopole ($\ell=0$) and quadrupole ($\ell = 2$), respectively.
Since the fiber assignment introduces the $\mu_k$-dependent anisotropy as shown in figure~\ref{fig:pkmu_real}, the observed quadrupole power spectrum becomes larger than the true power spectrum.
As in the case of the real-space power spectra, the IIP weighting recovers the true spectra up to $k \sim 0.2$ [$h$/Mpc].

\begin{figure}[]
	\begin{center}
		\includegraphics[width=1.0\textwidth]{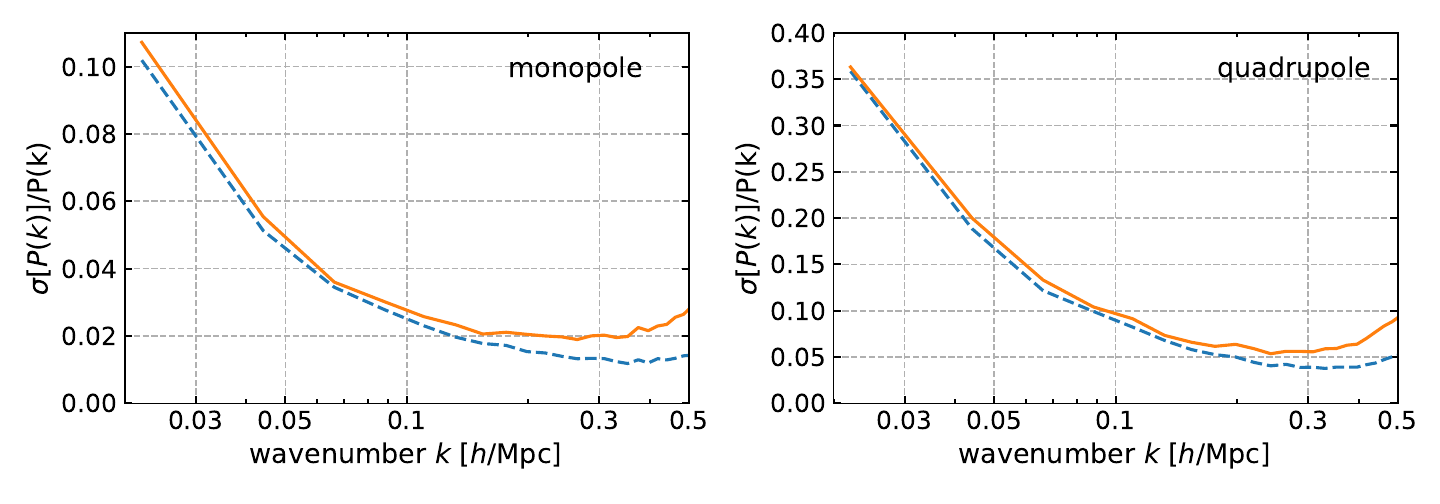}
		\caption{
		Fiber assignment effect on the fractional error of the galaxy power spectrum monopole (left panel) and the quadrupole (right panel) at $z = 1.3$.
		The blue solid lines are the fiducial simulation, where the fiber assignment is applied and corrected by the IIP weighting.
		The orange dashed lines show the fractional error of the galaxy power spectrum without the fiber assignment effects.
		}
		\label{fig:pk_gg_error}
	\end{center}
\end{figure}

As eq.~(\ref{eq:shot}) shows, the fiber assignment and the IIP weighting increase the shot noise, hence the fractional error of the power spectrum.
Figure~\ref{fig:pk_gg_error} shows the fractional error of the galaxy monopole and quadrupole at $z = 1.3$.
These relative errors are estimated from 500 realizations of the simulation.
We also show the fractional error of the power spectrum estimated from the simulation with the same galaxy number density with the ``observed'' sample but without the fiber assignment.
We find that the fiber assignment and the IIP weighting increase the fractional error by $\sim 5\%$ at $k = 0.05$ [$h$/Mpc] and $\sim 30\%$ at $k = 0.2$ [$h$/Mpc] compared to the case without the fiber assignment.

\subsection{Galaxy-lensing cross power spectrum}
We now investigate the fiber assignment effects on the galaxy-lensing cross power spectrum.
We measure the cross power spectrum as
\begin{equation}
C^{g\kappa}({\ell}) = 
\frac{1}{L^2}\left[\frac{1}{N_\ell}\sum_{\bm \ell}^{\ell \in \ell_b} \tilde{\delta}^{\rm 2D}_g({\bm \ell}) \tilde{\delta}_\kappa^*({\bm \ell}) \right],
\end{equation}
where $L$ is the side-length of the square-shape simulation field in radian, $N_\ell$ is the number of modes within a given multipole bin, and $\delta_\kappa$ is the Fourier transformed convergence field.
The projected galaxy density field $\delta^{\rm 2D}_g({\bm x})$ is obtained from the three-dimensional galaxy density field in the same manner as for the projection of matter density field, eq.(\ref{eq:projection}).

\begin{figure}[]
	\begin{center}
		\includegraphics[width=1.0\textwidth]{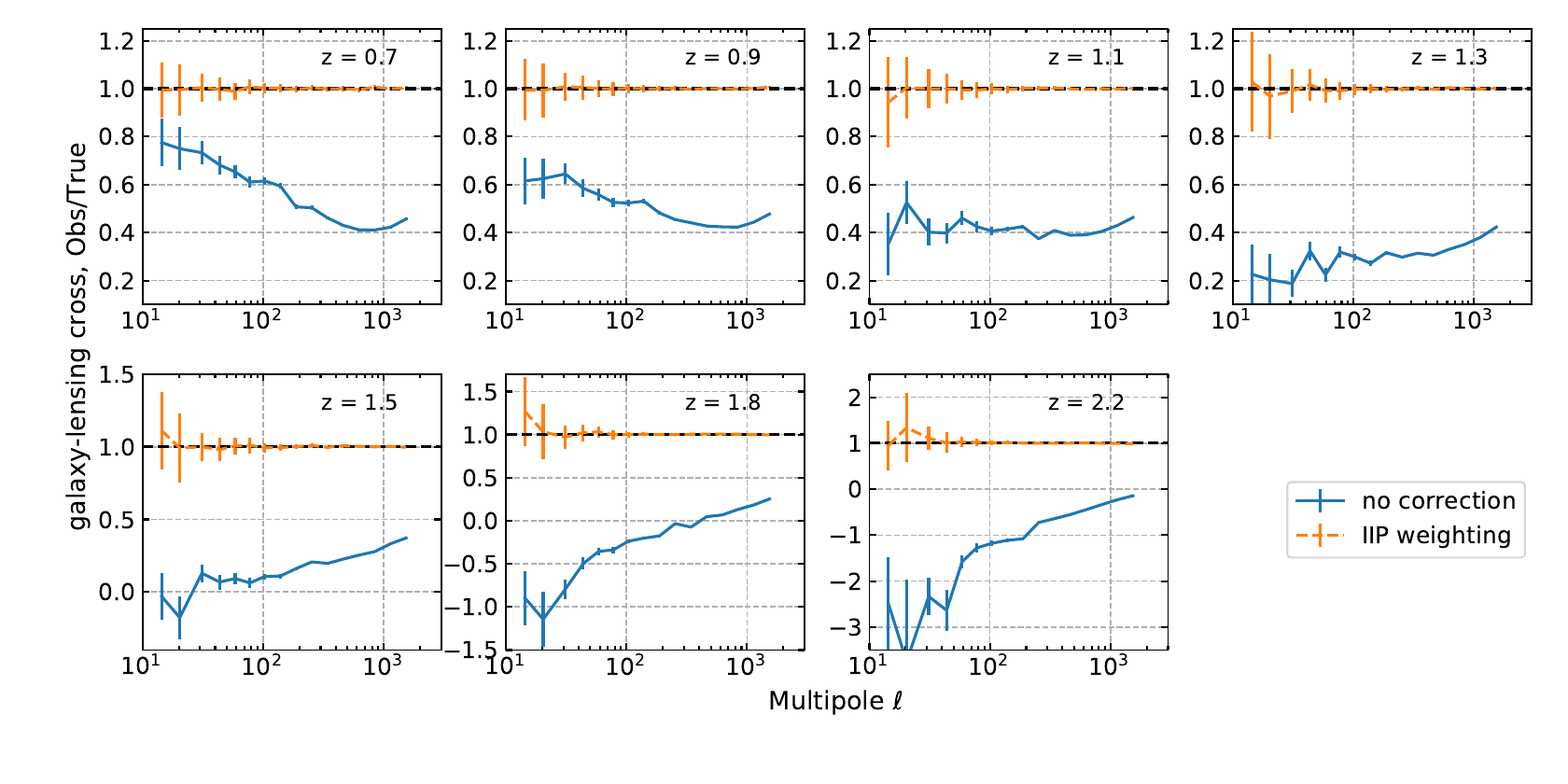}
		\caption{The ratio of the ``observed'' to ``true'' galaxy-lensing cross power spectrum for all redshift bins. 
		The blue solid lines show the results without any correction, while the orange dashed lines show that corrected by the IIP weighting. 
		}
		\label{fig:cl_gk_ets}
	\end{center}
\end{figure}

Figure~\ref{fig:cl_gk_ets} shows the ratio of the ``observed'' to ``true'' cross power spectra for all redshift bins, as well as those corrected by the IIP weighting.
We find that the observed cross power spectra are largely suppressed at all redshifts and even become negative at $z>1.8$.
Nevertheless, the IIP weighting recovers the true cross power spectra at an accuracy better than $1\%$ at all scales and redshift bins.
The strong suppression in the observed spectra is due to the correlation between the different redshift slices due to the fiber assignment.
Since the selection of one galaxy affects the selection of all the neighboring galaxies in angular space, the fiber assignment creates the artificial correlation of galaxies and underlying matter density fields between different redshift slices, which are not physically associated.
Since the weak lensing field is the integration of the matter density field along the LOS, this artificial correlation matters to the galaxy-lensing cross correlation.

\begin{figure}[]
	\begin{center}
		\includegraphics[width=1.0\textwidth]{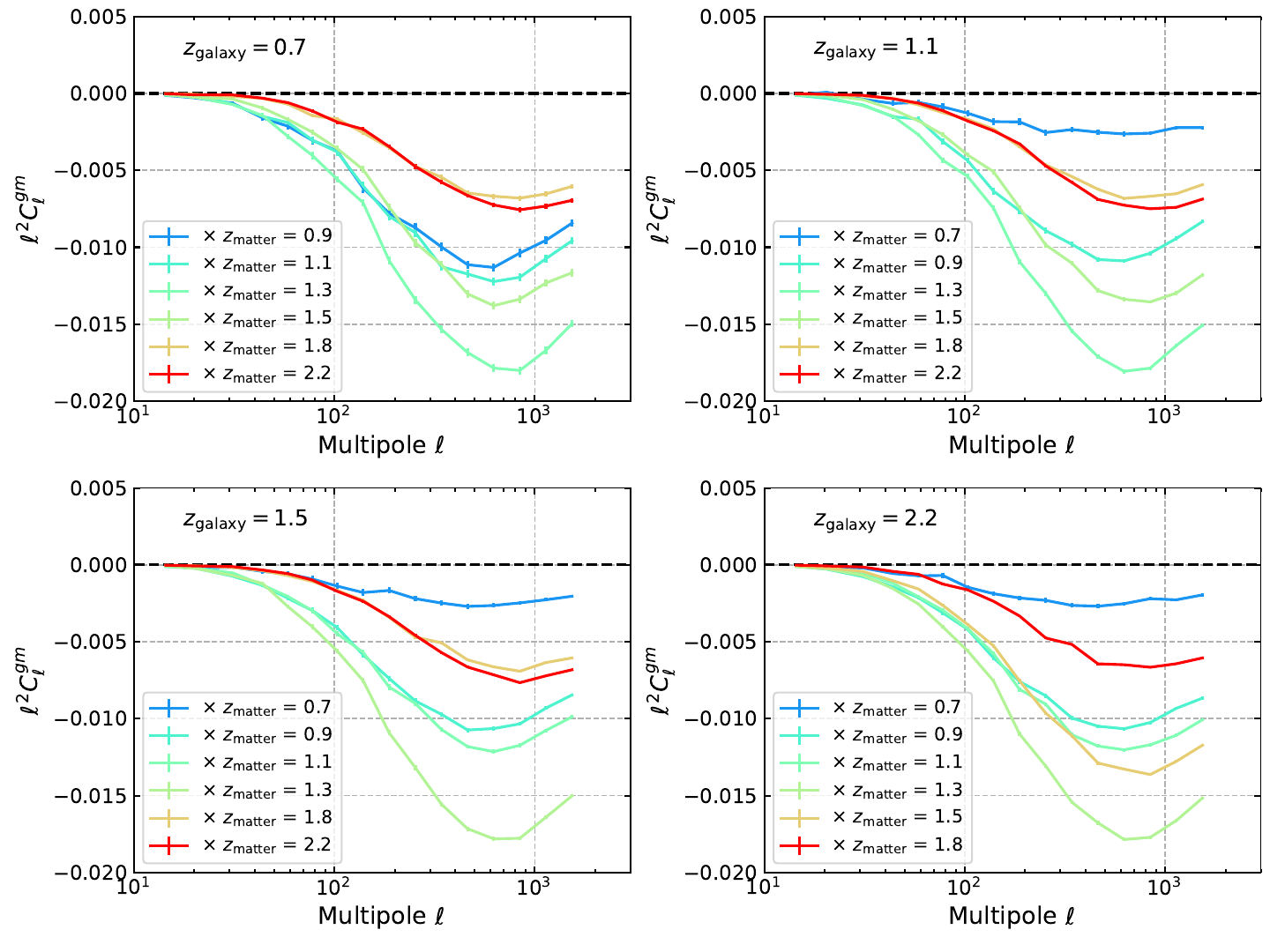}
		\caption{
		The cross power spectrum of the observed galaxy density field and the matter density field at the redshift different from galaxies.
		Data points show the mean of 500 realizations.
		The error of the mean is smaller than the size of the square data points. 
		We only show the results for the galaxies at $z = 0.7, 1.1, 1.5, 2.2$ for the clarity of the figure.
		Other redshifts show similar trends.
		}
		\label{fig:cl_gm_diffz}
	\end{center}
\end{figure}

\begin{figure}[]
	\begin{center}
		\includegraphics[width=1.0\textwidth]{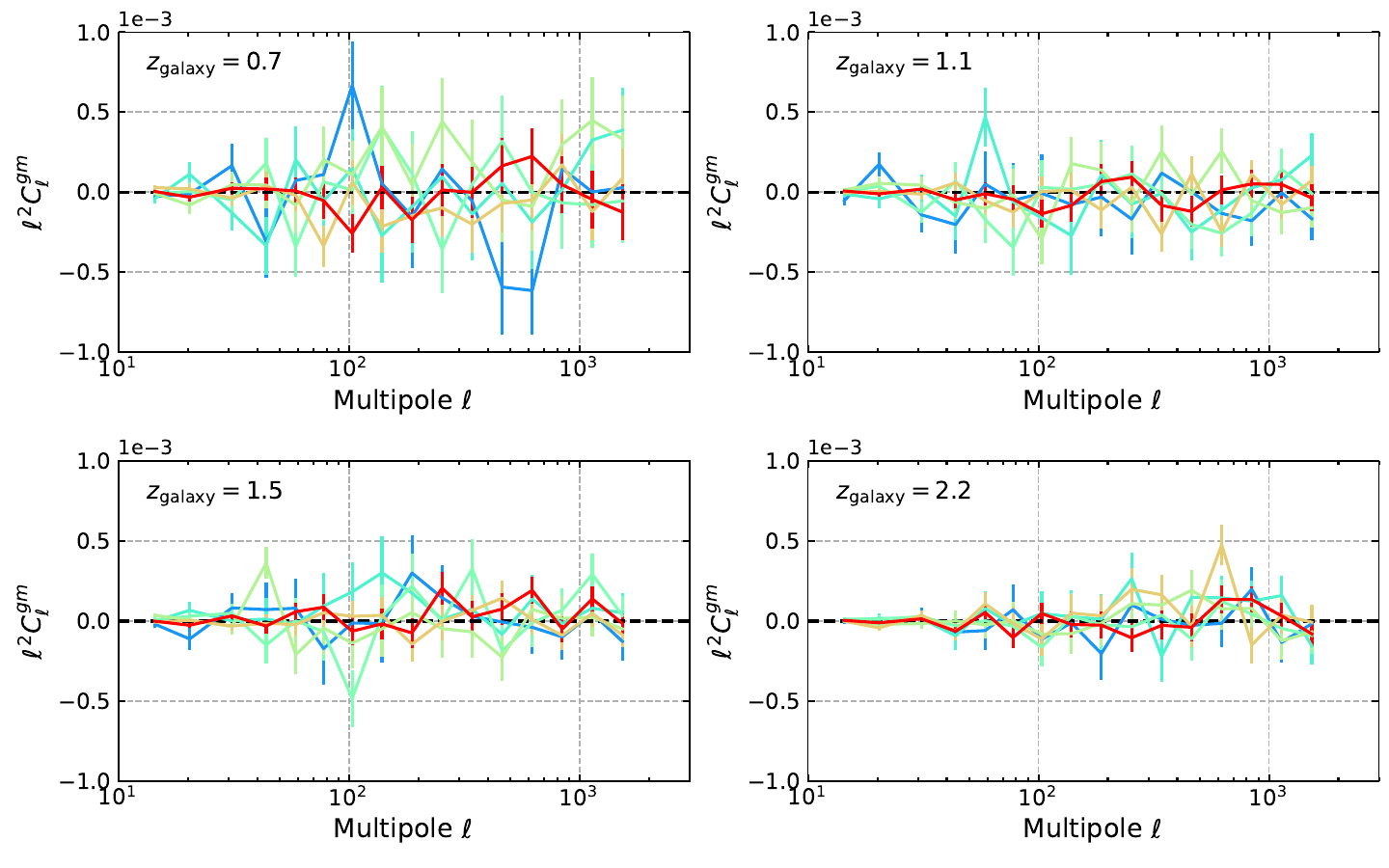}
		\caption{
		The same as figure~\ref{fig:cl_gm_diffz}, but for the results corrected by the IIP weighting.
		}
		\label{fig:cl_gm_ets_diffz}
	\end{center}
\end{figure}

To investigate this artificial correlation between galaxies and matter density fields at different redshifts, we compute the galaxy-matter cross power spectra.
Figure~\ref{fig:cl_gm_diffz} shows the projected galaxy-matter cross power spectra, where the matter density fields are at the different redshifts from galaxies.
We show the results with the galaxies at $z = 0.7, 1.1, 1.5,$ and $2.2$ as an example.
The negative matter-galaxy correlation is clearly appeared in the figure.
While the total amplitude of this artificial galaxy-matter anti-correlation does not depend on the redshift of galaxies, the amplitude of intrinsic galaxy-matter cross power spectrum is smaller at a higher redshift; thus the artificial anti-correlation dominates the observed galaxy-lensing cross power spectrum at high redshifts.
These negative signals are corrected by the IIP weighting, as shown in figure~\ref{fig:cl_gm_ets_diffz}.

\begin{figure}[]
	\begin{center}
		\includegraphics[width=1.0\textwidth]{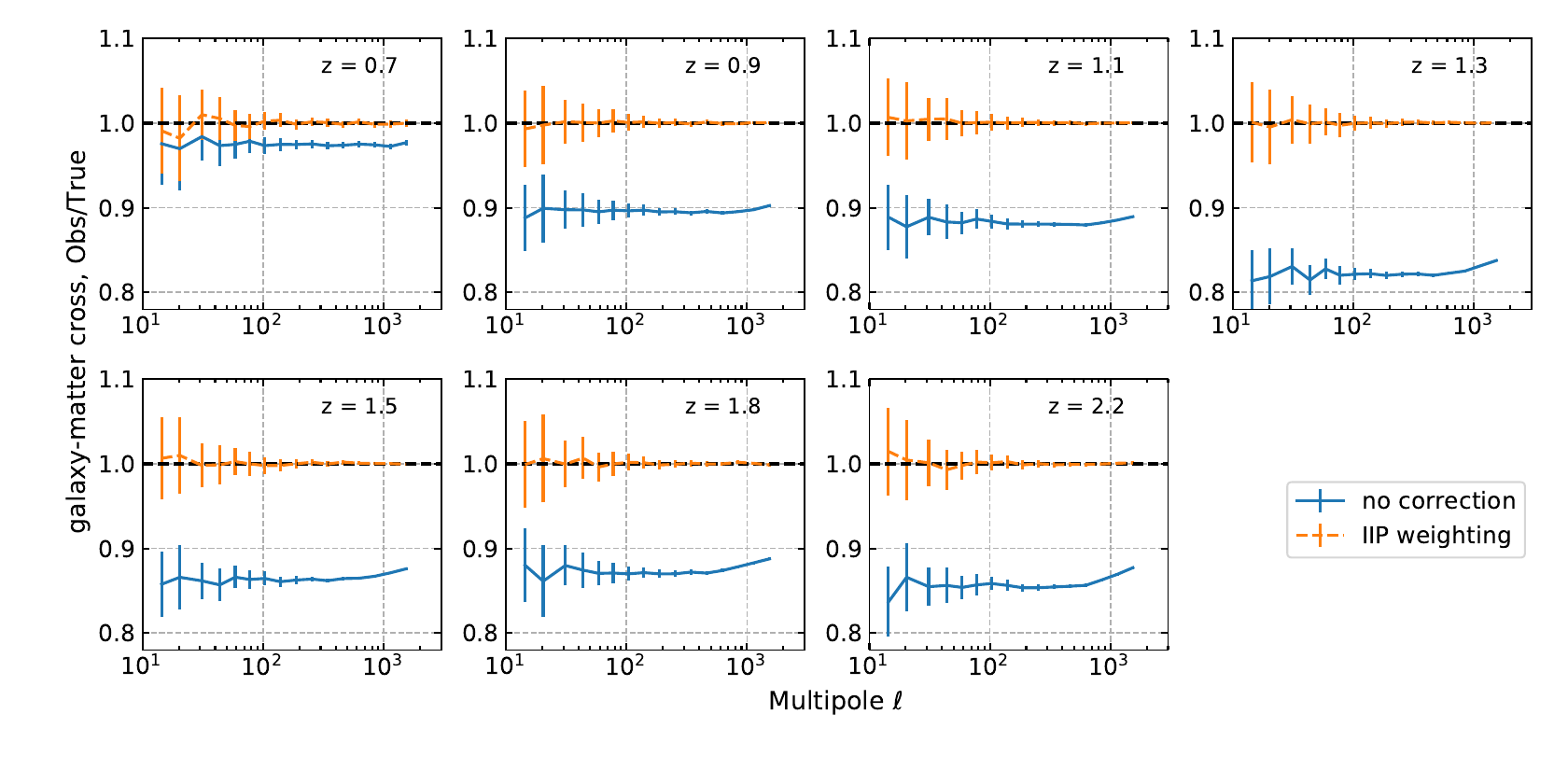}
		\caption{
        The ratio of the ``observed'' to ``true'' galaxy-matter cross power spectra for all redshift bins, where the matter density fields are at the same redshift with galaxies.
        The blue solid lines show the results without any correction, while the orange dashed lines show that corrected by the IIP weighting.
		}
		\label{fig:cl_gm_ets}
	\end{center}
\end{figure}

Figure~\ref{fig:cl_gm_ets} shows the ratio of the ``observed'' to ``true'' galaxy-matter cross power spectra for all redshift bins, where galaxies and matter density fields are at the same redshift.
The figure shows that the IIP weighting recovers the true spectrum up to the Nyquist frequency (the rightmost data points in the figure). This is not the case for the galaxy auto power spectra as discussed in section~\ref{subsec:power}.
This is because that the pairwise probability of galaxies does not enter in the estimator of galaxy-lensing cross power spectrum.

\subsection{Required accuracy of the IIP weights}
To estimate the IIP weights, we implicitly assume that all the galaxies in the ``true'' sample are our target galaxies.
In the real observation, however, non-target objects are included in the catalog of target galaxies.
This can systematically bias the IIP weights.
For example, if we misestimate the number of target galaxies in each tile, we also misestimates the weight for ``tiling effect'' as shown in eq.(\ref{eq:tile}) and demonstrated in ref.\cite{sunayama/etal:2020}.
This uncertainty in tiling weight is translated to the systematic shift of IIP weights at the spatial scale of each tile.
For the case of the HSC and PFS Cosmology program, this scale corresponds to $\sim 1\;{\rm deg^2}$.
To mimic this effect, we first divide the entire simulation FoV into the $1 \;{\rm deg^2}$ sub-regions.

Next, we add the random Gaussian fluctuations to the individual probability to be observed, $p_{i}$, while keeping the galaxies in the same sub-region have the same fluctuation.
We force $p_{i}$ does not exceed the natural range, $0 < p_{i} \leq 1.0$.

\begin{figure}[]
	\begin{center}
		\includegraphics[width=1.0\textwidth]{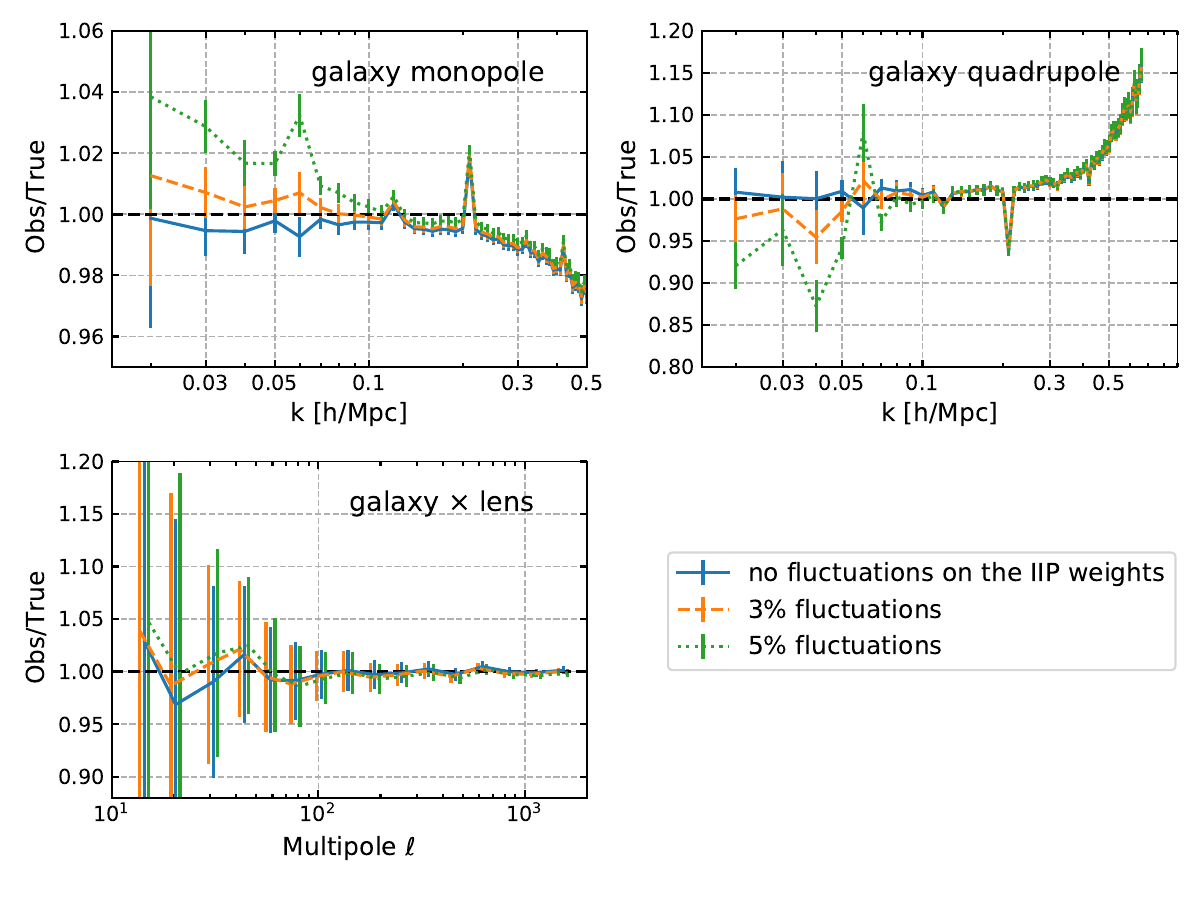}
		\caption{
		Effects of the systematic fluctuation in the IIP weights on the measurement of the galaxy power spectrum multipoles (upper panels) and the galaxy-lensing cross power spectrum (lower panel).
		The data points show the ratio of the ``observed'' to ``true'' power spectra for the galaxy sample at $z = 1.3$.
		The blue solid lines are the case without any systematic fluctuation in the IIP weights.
		The orange dashed lines (green dotted lines) show the case when we divide the galaxy sample into 1 ${\rm deg}^2$ grids and add the 3\% (5\%) Gaussian fluctuations keeping that the galaxies in the same grid have the same fluctuation.
		}
		\label{fig:error_iip}
	\end{center}
\end{figure}

The upper panels of figure~\ref{fig:error_iip} show the ratio of the ``observed'' to ``true'' galaxy power spectra at $z = 1.3$, varying the fluctuation on the IIP weights.
We find that the 3\% (5\%) fluctuations in the IIP weights with 1~deg scale is translated to $\sim$1\% ($\sim$4\%) and $\sim$3\% ($\sim$10\%) systematic offsets in the monopole and quadrupole of power spectrum at large scales, respectively.
With a 3\% error of the IIP weights, the recovered galaxy power spectra are consistent with the true spectra within the statistical errors.
This result indicates that the imaging survey used for the target selection is required to achieve $<3\%$ precision in target number density to correct for the galaxy power spectra by the IIP weighting.
For the galaxy-lensing cross power spectrum, shown in the lower panel of figure~\ref{fig:error_iip}, the systematic offsets are negligible even with the 5\% fluctuations in the IIP weights.
This is because that the correlation of the IIP weights, which causes the systematic offsets in the auto power spectra, does not enter in the cross power spectra.

\section{Summary and Conclusions}
\label{sec:summary}
In this paper, we have investigated the fiber assignment effect on the galaxy power spectrum and the galaxy-lensing cross power spectrum.
We have shown that the fiber assignment suppresses the amplitude of galaxy power spectrum at all scales, as in the case of the galaxy clustering in configuration space \cite{sunayama/etal:2020}.
We newly found that the amplitude of the galaxy-lensing cross power spectrum is also suppressed by the fiber assignment, which even become negative at high redshifts.
This is because the fiber assignment introduces the artificial anti-correlation of galaxies and matter density fields at different redshifts.

To mitigate the fiber assignment effect, we have tested the weighting method using the inverse probability of the galaxies to be observed. 
The weighting method recovers the galaxy power spectrum multipoles better than $\sim 1 \%$ accuracy up to $k \sim 0.2$ [$h$/Mpc], but we need to consider the pairwise probability of galaxy pairs to be observed to correct for the galaxy power spectrum at $k > 0.2$ [$h$/Mpc].
The galaxy-lensing cross power spectrum is recovered better than $\sim 1 \%$ up to the Nyquist frequency by the IIP weighting.
This is because the pairwise probability does not enter in this statistics.

As we demonstrated in this paper, the galaxy power spectrum and the galaxy-lensing cross power spectrum have the different responses to the fiber assignment and the different correction methods are required for these measurements.
To maximize the scientific outcomes of the photometric and spectroscopic surveys, the joint analysis of the power spectrum and the galaxy-lensing cross power spectrum is crucial; thus, it is important to understand how the fiber assignment affects these measurements and how the effects are mitigated.

\acknowledgments
We thank Eiichiro Komatsu and Shun Saito for their careful comments., and thank the PFS collaboration, especially the cosmology working group, for discussions. 
We also thank Atsushi Shimono and Society of Photo-Optical Instrumentation
Engineers for permission to reuse the figure.
RM is supported by JSPS KAKENHI Grant Number 20K14515. 
RM also thanks the Ministry of Science and Technology (MOST) for support through grant
MOST 108-2112-M-001-007-MY3, the Academia Sinica for Investigator Award AS-IA-109-M02.
TS is supported by Grant-in-Aid for JSPS Fellows 20J01600 and JSPS KAKENHI Grant Number 20H05855.

\bibliography{references}
\end{document}